\documentclass[fleqn,usenatbib]{mnras}

\usepackage{newtxtext,newtxmath}

\usepackage[T1]{fontenc}
\usepackage{ae,aecompl}

\usepackage{xspace} 
\newcommand{\sage}{\texttt{SAGE}\xspace}
\newcommand{\dustysage}{\texttt{Dusty~SAGE}\xspace} 

\newcommand{\Msun}{\ensuremath{\mathrm{M}_{\odot}}\xspace}

\usepackage{accents} 
\newcommand*{\dt}[1]{%
  \accentset{\mbox{\large\bfseries .}}{#1}}

\usepackage{graphicx}	
\usepackage{amsmath}	
\usepackage{xcolor}     

\usepackage{comment}

\usepackage[normalem]{ulem} 

\hypersetup{breaklinks=true}

\title[Exploring the relation between dust mass and galaxy properties using \dustysage]{Exploring the relation between dust mass and galaxy properties using \dustysage}

\author[D. P. Triani et al.]{
Dian P. Triani,$^{1,2}$\thanks{E-mail: dtriani@swin.edu.au}
Manodeep Sinha,$^{1,2}$
Darren J. Croton,$^{1,2}$
Eli Dwek,$^{3}$
\newauthor
and Camilla Pacifici$^{4}$
\\
$^{1}$Centre for Astrophysics \& Supercomputing, Swinburne University of Technology, Hawthorn, VIC 3122, Australia\\
$^{2}$ARC Centre of Excellence for All Sky Astrophysics in 3 Dimensions (ASTRO 3D)\\
$^{3}$Observational Cosmology Lab, NASA Goddard Space Flight Center, Code 665, Greenbelt, MD 20771, USA\\
$^{4}$Space Telescope Science Institute, Baltimore, MD 21218, USA\\
}

\date{Accepted XXX. Received YYY; in original form ZZZ}

\pubyear{2021}

\begin{document}
\label{firstpage}
\pagerange{\pageref{firstpage}--\pageref{lastpage}}
\maketitle

\begin{abstract}
We explore the relation between dust and several fundamental properties of simulated galaxies using the \dustysage semi-analytic model. In addition to tracing the standard galaxy properties, \dustysage also tracks cold dust mass in the interstellar medium (ISM), hot dust mass in the halo and dust mass ejected by feedback activity. Based on their ISM dust content, we divide our galaxies into two categories: ISM dust-poor and ISM dust-rich. We split the ISM dust-poor group into two subgroups: halo dust-rich and dust-poor (the latter contains galaxies that lack dust in both the ISM and halo). Halo dust-rich galaxies have high outflow rates of heated gas and dust and are more massive. We divide ISM dust-rich galaxies based on their specific star formation rate (sSFR) into star-forming and quenched subgroups. At redshift $z=0$, we find that ISM dust-rich galaxies have a relatively high sSFR, low bulge-to-total (BTT) mass ratio, and high gas metallicity. The high sSFR of ISM dust-rich galaxies allows them to produce dust in the stellar ejecta. Their metal-rich ISM enables dust growth via grain accretion. The opposite is seen in the ISM dust-poor group. Furthermore, ISM dust-rich galaxies are typically late-types, while ISM dust-poor galaxies resemble the early-type population, and we show how their ISM content evolves from being dust-rich to dust-poor. Finally, we investigate dust production from $z=3$ to $z=0$ and find that all groups evolve similarly, except for the quenched ISM dust-rich group.
\end{abstract}




\section{Introduction}
\label{sec:intro}
The cosmic star formation history shows a rise of star formation activity in our Universe from high-redshift till redshift $z=2$, known as cosmic noon, and then a decrease towards $z=0$ \citep{Madau14}. Our current understanding is that star formation depends on the availability of $\mathrm{H_2}$ clouds, the raw material for stars \citep{Kennicutt12}. As a galaxy evolves, $\mathrm{H_2}$ clouds are consumed to form stars, and less material is available for the next star formation episode. However, infalling gas can provide a fresh supply which replenishes gas that was consumed by star formation activity \citep[e.g][]{Dekel06, Almeida14}.

Morphology and colour often correlate with star formation in galaxies \citep[e.g.][for a review]{RH94}. Elliptical galaxies are dominated by an old stellar population and appear red \citep{Thomas05, Kormendy09}. These elliptical galaxies have run out their star-forming gas; therefore, they have no or low star formation. At the opposite end, spiral galaxies have bluer emission, showing signs for young stellar populations and active star formation \citep[see also a review from][]{Kennicutt98}.

Per ``conventional'' galaxy evolution theory, galaxies evolve from star-forming spiral galaxies to quenched ellipticals \citep{Bundy06, Skelton12, Tojeiro13}. Dust follows this evolutionary sequence, and galaxies become dust-poor at late-times, either through dust destruction or ejection out of the ISM. Galaxy formation models assume that as star formation activity decreases, mergers and disk instabilities make galaxies more spheroidal \citep{Somerville01, Baugh05, Croton06, Croton16}. During such morphological transformations, a fraction of the ISM dust is ejected through SN and AGN feedback. Therefore, we see a decrease in the ISM dust mass in more elliptical galaxies.

Since the dust content of a galaxy depends on its unique evolutionary history, the observed dust content in the local and high-redshift Universe provides insight into the physics of galaxy formation. Studies of high-redshift galaxies in the far-infrared and submillimetre regime reveal an abundance of star-forming galaxies with massive dust reservoirs \citep{Valiante09}. Dusty galaxies are even found in the very early Universe, including A1689-zD1 at $z=7.5$ with a dust mass of $4 \times 10^7$ \Msun \citep{Watson15}, and HFSL3 at $z=6.34$ with a dust mass of $1.3 \times 10^9$ \Msun \citep{Riechers13}. However, observations also see star-forming galaxies at early and late times with little dust and low metallicity \citep{Fisher14}. The dichotomy between dust-rich and dust-poor galaxies across redshift motivates this work.
 
Recent galaxy surveys have measured the dust mass function \citep{DEE03, VDE05, Dunne11, Eales09, Clemens13} and many scaling relations between dust and the fundamental properties of galaxies. These include the relationship between dust mass and stellar mass \citep{Santini14} and the relationship between dust mass and star formation rate (SFR) \citep{daCunha10, Santini14}. These scaling relations provide constraints for galaxy evolution models that track the dust properties of galaxies.

Historically, dust in galaxies is commonly investigated using one zone analytical models \citep[e.g.,][]{Dwek1998, ZGT08, Valiante09, Asano13, Zhukovska14}. Such models are critical to assess the various processes of dust production and destruction, especially to reproduce the dust content of specific galaxy populations. However, they generally use a simplistic approach to model gas and stellar evolution in galaxies and do not provide predictions for galaxies within a cosmological volume. Zoom-simulations coupled with self-consistent dust tracking give a more realistic accounting of how dust, metals, gas and stellar populations interact within galaxy evolution framework \citep{Bekki15, Aoyama17}. But they are computationally expensive and tend to focus on reproducing the properties of individual galaxies. 

\cite{McKinnon16} was one of the first works to model dust and galaxy co-evolution within a cosmological simulation. They incorporated a detailed dust prescription to a hydrodynamical simulation, including stellar production, grain growth and grain destruction by supernovae shocks. To date, a few semi-analytic galaxy models have included self-consistent dust modelling. These models can reproduce a number of global trends between dust and various galaxy properties across cosmic time, with less detail but more efficient computing cost and time compared to the hydrodynamical simulations \citep{Popping17, Vijayan19, Triani20}.

In this paper, we extend the analysis of \citet{Triani20} to investigate the different characteristics between dust-rich and dust-poor simulated galaxies using the \dustysage\footnote{\url{https://github.com/dptriani/dusty-sage}} semi-analytic model. Dust in the ISM forms in stellar ejecta \citep{Dwek1998, ZGT08}, grows further via grain accretion \citep{Draine90, Dwek1998, ZGT08, Draine09}, and is destroyed by supernovae shocks \citep{DS80, Jones94, ZGT08, Slavin15}. \dustysage includes analytical prescriptions for these mechanisms, as well as dust locked in stars (astration), and dust inflows from and outflows to the halo and ejected reservoirs. In the model, dust undergoes further destruction via thermal sputtering in the halo and ejected reservoirs due to their high temperature. The primary constraints for dust modelled by \dustysage are the observed dust mass function and dust mass - stellar mass relation at $z=0$. \dustysage includes dust tracking in the ISM, halo, and ejected dust by feedback processes. It provides predictions for the dust mass function and the relation between dust mass and stellar mass at high redshift.

This paper is organised as follows: In Section \ref{sec:dustysage}, we provide a brief description of \dustysage. We study how the dust fraction in each reservoir relates to the stellar mass and morphology in Section \ref{sec:morphologies}. Then we categorise galaxies into groups based on their ISM and halo dust content in Section \ref{sec:classification}. In Section \ref{sec:relation}, we map these groups based on their morphology, star formation activity and stellar mass at redshift $z=0$. We extend the relations to higher redshift, up to $z=3$, in Section \ref{sec:evolution}. Section \ref{sec:discussion} and \ref{sec:conclusion} provide a discussion and conclusion, respectively. Throughout this paper, we assume $h = 0.73$ based on the cosmology used of the Millennium simulation.

\section{The semi-analytical model - Dusty SAGE}
\label{sec:dustysage}

Here we provide only a brief overview of the galaxy and dust model used in this work. \dustysage is built on the more generic galaxy model \sage \citep{Croton06, Croton16}, but with a detailed prescription that tracks dust evolution. The summary of both galaxy and dust evolution mechanisms in \texttt{Dusty SAGE} is presented in Figure 1 in \citet{Triani20}. The model follows baryonic growth in dark matter halos taken from an N-body simulation. There are four baryonic reservoirs: the pristine gas, the hot halo, the ISM and an ejected reservoir. In this paper, we run \dustysage on the Millennium simulation \citep{Springel05} and select galaxies  with stellar mass $\log M_* = 9 - 12$ \Msun. 

Mass is exchanged among the different baryonic reservoirs across cosmic time. Pristine gas falls into the collapsed dark matter halo and is heated to the virial temperature -- the hot halo reservoir \citep{RO77, WR78}. A fraction of this hot gas then cools into the ISM \citep{SD93}. In the ISM, cold hydrogen differentiates into atomic and molecular hydrogen \citep{HMK79}. Stars then form from molecular clouds \citep{Kennicutt12}. During their evolution, stars create helium and elements heavier than helium in their core, known as `metals'. These metals are expelled in stellar ejecta and change the metallicity of the ISM (ratio of metals and gas mass). Massive stars end their life as supernovae and inject a large amount of energy into the cold ISM. This energy reheats cold ISM gas back to the hot halo reservoir, and potentially even expels it outside the halo -- the `ejected' reservoir \citep[e.g.,][]{MW03}. This process is known as SN feedback. The gas accretion onto supermassive black holes provides another source of feedback; this energetic phenomenon is known as active galactic nuclei (AGN) \citep{Croton06}. The ejected gas can later be reincorporated back to the halo and galaxy.

In both \dustysage and \sage \citep{Croton06, Croton16}, the bulge-to-total stellar mass ratio represents the morphology of the galaxy, which evolves during disk instabilities and mergers. We adopt the disk stability criteria of \citet{1998MMW} and transfer sufficient stellar mass from the disk to the bulge to ensure stability every time an instability occurs. In a merger between a more massive central galaxy and a less massive satellite galaxy, bulge enrichment depends on the total stellar and gas mass ratio of both progenitors. If the mass ratio exceeds 0.3, a `major' merger has occurred: the disks of both galaxies are destroyed and all stars are placed in a bulge. Otherwise, the merger is `minor' and the satellite stellar content is added to the central bulge. Both instabilities and mergers can trigger a starburst. Unlike quiescent disk star formation, stars formed in a merger-induced starburst are placed in the bulge.

Besides the usual galaxy evolution processes, \dustysage also incorporates a detailed dust evolution model. Analogous to the gas, we track dust in three reservoirs -- the ISM, the hot halo and an ejected reservoir. Dust is mainly processed in the ISM, then heated to the halo and ejected reservoirs via feedback mechanisms powered by supernovae and AGN. The total dust production rate in the ISM is described by:
\begin{equation} \label{eq:total_rate}
    \dt{M}_\mathrm{d} = \dt{M}_\mathrm{d}^\mathrm{form} + \dt{M}_\mathrm{d}^\mathrm{growth} - \dt{M}_\mathrm{d}^\mathrm{dest} - \dt{M}_\mathrm{d}^\mathrm{SF} - \dt{M}_\mathrm{d}^\mathrm{outflow} + \dt{M}_\mathrm{d}^\mathrm{inflow},
\end{equation}
where:
\begin{itemize}
    \item $\dt{M}_\mathrm{d}^\mathrm{form}$ is the stellar dust formation rate: In every star formation episode, \dustysage tracks the abundance of C, N and O in AGB stars and C, O, Mg, Si, S, Ca and Fe in SN II ejecta. The condensation of these elements to form dust is given in Table \ref{tab:params}. 
    \item $\dt{M}_\mathrm{d}^\mathrm{growth}$ is the grain growth rate in dense molecular clouds: Existing grains grow via metal accretion \citep{Dwek1998, ZGT08} where the timescale for this process depends on the metal abundance in the cold gas.
    \item $\dt{M}_\mathrm{d}^\mathrm{dest}$ is the destruction rate via SN shocks: We follow the prescription from \cite{DS80, Mckee89, Asano13} to compute a destruction timescale from the total cold gas mass and the supernovae rate, efficiency and swept mass.
    \item $\dt{M}_\mathrm{d}^\mathrm{SF}$ is the rate for dust locked in newly formed stars, which is proportional to the star formation rate and the dust-to-gas ratio.
    \item $\dt{M}_\mathrm{d}^\mathrm{outflow}$ and $\dt{M}_\mathrm{d}^\mathrm{inflow}$ are the dust outflow and inflow rates. SN and AGN feedback can reheat cold ISM gas and expel it to the halo. The feedback energy, if large enough relative to the depth of the potential well, can even eject the gas to leave the galaxy and host halo. In an outflow, we assume that the dust-to-gas ratio of the ejected gas is equal to the ISM. Ejected gas can be reincorporated back to the halo, while maintaining the dust-to-gas ratio of the ejected reservoir. Hot gas undergoes cooling process back to the disk, and we assume in this inflow the dust-to-gas ratio equals that of the halo. 
\end{itemize}

Dust populates the halo and ejected reservoirs through outflows from the ISM. In both reservoirs, dust grains are destroyed via thermal sputtering on a short timescale that depends on the gas density and temperature. In both the halo and ejected reservoirs, we assume the virial temperature, with an isothermal density profile for the hot gas extending to the virial radius, and a uniform density profile for the ejected component. In every timestep, the ejected reservoir's density is evaluated by dividing the gas mass with the reservoir's volume, assuming the virial radius. However, the computed timescale should be taken as the upper limit for the ejected reservoir. In reality, the ejected reservoir is a mix of the circumgalactic medium (CGM) and intergalactic medium (IGM). Gas in the IGM might extend beyond the virial radius, resulting in a lower density. The temperature might also be higher than in the model, allowing for a more efficient sputtering. 

Although thermal sputtering is very efficient, we still find a significant abundance of dust in the halo. Dust properties in the halo depend on the balance between the outflows and inflows from/to the ISM. The depth of the galaxy's potential well also affects the outflow mass trapped in the halo. Galaxies with a shallow potential are more likely to lose the outflowing materials to the ejected reservoir. We assume that the gas outflow carries dust in the same proportion as the ISM and transfers it to the halo, with no dust destruction in the process. It may well be that a fraction of dust is destroyed when ejected out of the ISM, or the DTG ratio of the outflow differs from the ISM, which will alter our predictions. As we mentioned above, it is also possible that the sputtering rates are higher than our prediction. However, the nature of such processes is currently unclear. Future observations to quantify dust properties in the galactic wind and the IGM will provide additional constraints to these processes. 

Figure 14 in \citet{Triani20} shows that galaxies have more dust in their halo than in the ISM at low-redshift. This outcome roughly agrees with the massive dust content found in the CGM \citep{Dunne11, PMC15}. The galaxy formation model of \citet{Popping17} also predicts a significant amount of dust in both the halo and ejected reservoir. However, their dust mass density is notably higher than our results.
 
\dustysage provides a good agreement with the galaxy stellar mass function at redshift $z=0$. It also successfully reproduces the dust mass function and various dust scaling relations over a wide range of redshifts \citep{Triani20}. To achieve a more realistic distribution of the stellar mass in the bulge and disc, we make slight changes to a few of the parameters in \citet{Triani20}. These are listed in Table \ref{tab:params}. 

\begin{table*}
 \caption{Fiducial \dustysage parameters used throughout this work, also compared to those from \citet{Triani20}.}
 \label{tab:params}
 \begin{tabular}{lccc}
  \hline
  \hline
  Parameter & Description & Value &\citet{Triani20} \\
  \hline
  \hline
  $f_b^\mathrm{cosmic}$ & Cosmic baryon fraction & 0.17 & 0.17\\
  \hline
  $z_0$ & Redshift when $\mathrm{H_{II}}$ regions overlap & 8.0 & 8.0\\
  $z_r$ & Redshift when the intergalactic medium is fully reionized & 7.0 & 7.0\\
  \hline
  $\alpha_\mathrm{SF}$ & Star formation efficiency from $\mathrm{H_2}$ [$\mathrm{Myr^{-1}}$] & 0.005 & 0.005\\
  $R$ & Instanteneous recycling fraction & 0.43 & 0.43\\
  \hline
  $\epsilon_\mathrm{disc}$ & Mass-loading factor due to supernovae & 2.0 & 3.0\\
  $\epsilon_\mathrm{halo}$ & Efficiency of supernovae to unbind gas from the hot halo & 0.2 & 0.3\\
  $k_\mathrm{reinc}$ & Velocity scale for gas reincorporation & 0.15 & 0.15\\
  \hline
  $\kappa_\mathrm{R}$ & Radio mode feedback efficiency & 0.09 & 0.08\\
  $\kappa_\mathrm{Q}$ & Quasar mode feedback efficiency & 0.005 & 0.005\\
  $f_\mathrm{BH}$ & Rate of black hole growth during quasar mode & 0.015 & 0.015\\
  \hline
  $f_\mathrm{friction}$ & Threshold subhalo-to-baryonic mass for satellite disruption or merging & 1.0 & 1.0\\
  $f_\mathrm{major}$ & Threshold mass ratio for merger to be major & 0.15 & 0.3\\
  \hline
  $\alpha_\mathrm{burst}$ & Exponent for the powerlaw for starburst fraction in merger & 0.18* & 0.7\\
  $\beta_\mathrm{burst}$ & Coefficient for starburst fraction in merger & 0.75* & 0.56\\
  \hline
  $\delta_\mathrm{C}^\mathrm{AGB}$ & Condensation efficiency for AGB stars & 0.2 & 0.2\\
  $\delta_\mathrm{C}^\mathrm{SNII}$ & Condensation efficiency for  SN II & 0.15 & 0.15\\
  $\tau_\mathrm{acc,0}$ & Accretion timescale for grain growth [$\mathrm{yr}$] & $4.0 \times 10^5$ & $4.0 \times 10^5$\\
  $f_\mathrm{SN}$ & Fraction of destroyed dust to the swept dust mass by SN & 0.1 & 0.1\\
  \hline
  \hline
  \multicolumn{3}{l}{*value adopted from \citet{Somerville01}}\\
 \end{tabular}
\end{table*}

\section{How morphology and stellar mass relate to dust fraction in the hot halo, ISM and ejected reservoir}
\label{sec:morphologies}

\begin{figure*}
    \centering
    \includegraphics[width=0.8\textwidth]{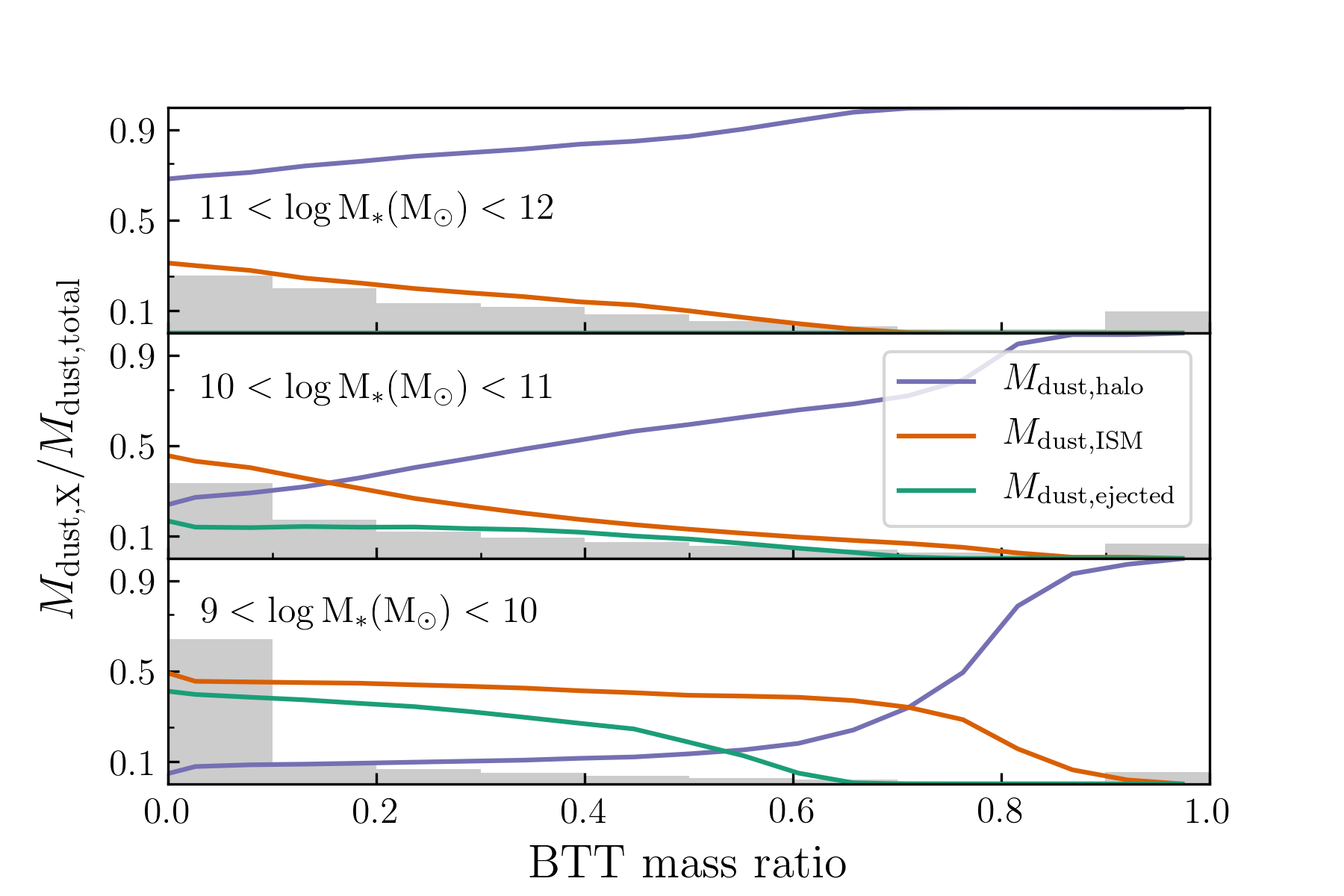}
    \caption{The median value for dust mass fraction in each baryon reservoir versus bulge-to-total stellar mass ratio (BTT). The purple, orange and green lines represents hot dust mass in the halo, cold dust mass in the ISM, and dust in the ejected reservoir, respectively. The grey histogram represents the number distribution of galaxies in 10 BTT bins.}
    \label{fig:btt-fraction}
\end{figure*}

In this section, we investigate the relation between galaxy morphology, represented as the bulge-to-total (BTT) mass ratio, and the fraction of dust in the ISM, halo and ejected reservoirs relative to the total dust mass in all reservoirs. We use stellar mass in computing the BTT mass ratio and assume the gas mass to be negligible. We present the median dust fraction vs BTT mass ratio in three stellar mass bins in Figure \ref{fig:btt-fraction}. The grey histogram shows the distribution of BTT mass ratios across all galaxies. In all panels, galaxies with higher BTT mass ratio contain most of their dust in the halo while those with a lower BTT mass ratio tend to keep theirs in the ISM. For the low mass galaxies in the bottom panel, spiral galaxies contain a significant fraction of their dust in the ISM and ejected reservoir. These galaxies make up more than 50\% of the overall population. Conversely, the halo contains nearly 90\% of the total dust in elliptical galaxies with BTT > $0.8$.

For the galaxies with $\log \mathrm{M_*}\ (\Msun) = 10 - 11$ (central panel), the same trend occurs. Galaxies with a BTT mass ratio smaller than $\sim 0.2$ keep almost half of their dust in the ISM. For the most massive galaxies (top panel), all galaxies contain most of their dust in the halo. Massive galaxies are more likely to host AGNs compared to low-mass galaxies. The stronger AGN feedback is more likely to blow dust out into the halo (and ejected reservoir) compared to the weaker SN feedback commonly found in low-mass galaxies. Massive galaxies also have deeper potential wells than the low mass systems, allowing them to capture reheated dust in the hot halo. However, although the ISM dust fraction is smaller than in Milky Way sized galaxies, their actual dust mass in the ISM is significant and will likely make a noticeable contribution to the galaxy SED in the infrared.

Observations have found galaxies with a massive amount of dust in their CGM. For a galaxy with stellar mass $\log \mathrm{M_*}\ (\Msun) = 10.1$, \citet{Peeples14} found a fiducial value of metals locked in CGM dust of $5 \times 10^7$ \Msun, almost twice the value for those found in the form of ISM dust, $2.6 \times 10^7$ \Msun. \citet{Menard10} derived a CGM dust mass of $5 \times 10^7$ \Msun for galaxies with stellar mass $\log \mathrm{M_*} = 10.4 \Msun$. Although the CGM is not exactly equivalent to our halo gas component, we will use CGM dust as a proxy to test our predictions of the dust bond to, but outside the galaxy.

\section{Classifying galaxies based on their dust and fundamental properties}
\label{sec:classification}

To investigate how the dust content of galaxies correlates with their stellar mass, star formation activity, and morphology, we first establish our simulated galaxies into groups. In the \dustysage model, the dust in galaxies is distributed in three components - the ISM, the halo and the ejected reservoir. We exclude the dust in the ejected reservoir in our analysis since it is out of the system. Our four dust groups are described below.

\begin{figure}
    \centering
    \includegraphics[width=0.5\textwidth]{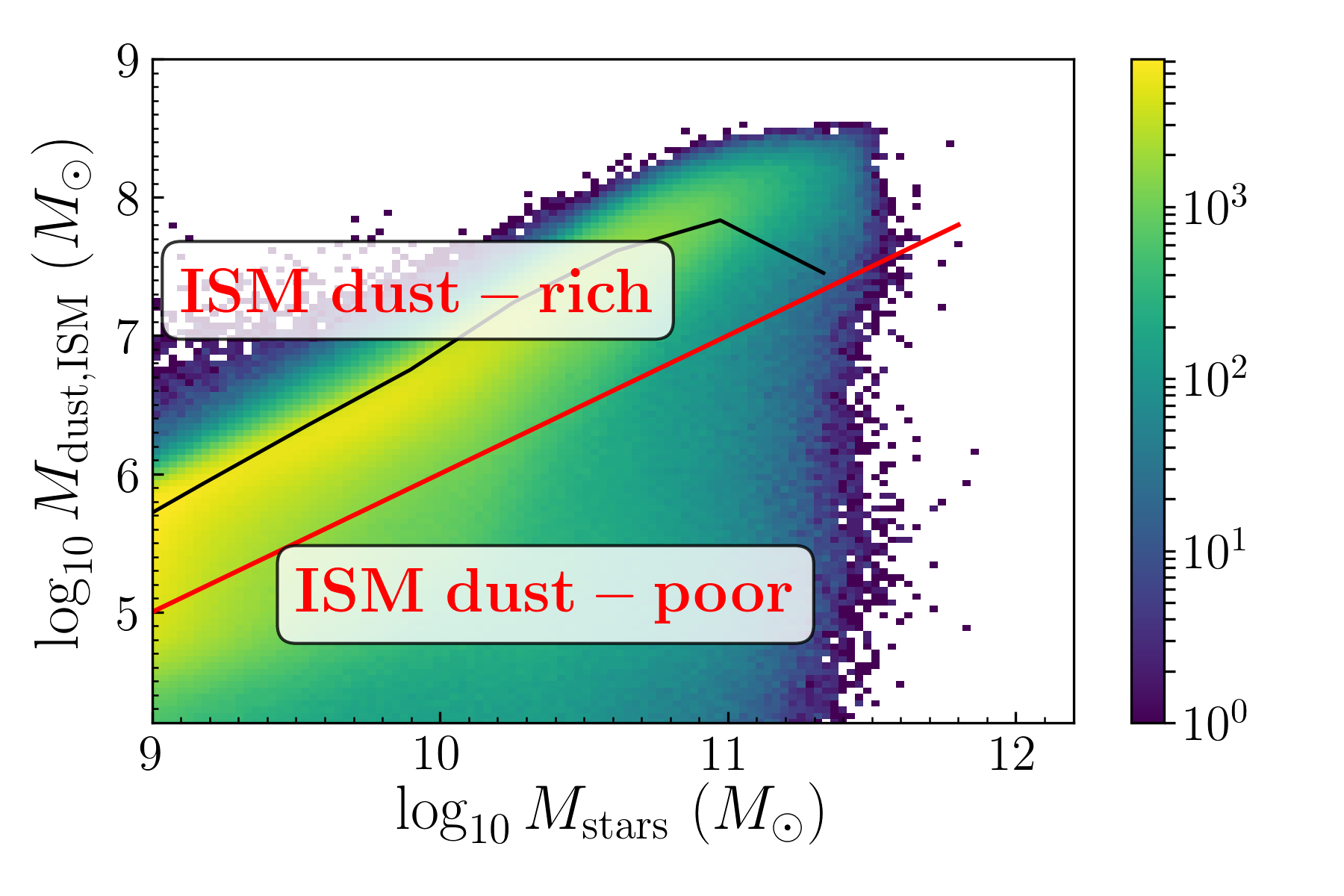}
    \caption{The cold dust mass of simulated galaxies as a function of stellar mass at redshift $z=0$. The heatmap shows the density distribution with a brighter colour representing higher density, and the black line marks the median. The red line marks the threshold where the cold dust mass per stellar mass equals $10^{-4}$. Above this line, galaxies are classified as ISM dust-rich while below are classified as ISM dust-poor.}
    \label{fig:dust-stellarmass}
\end{figure}

\subsection{ISM dust-poor}
\label{ssec:ISM-dust-poor}

When considering dust, observers typically report the properties of the ISM \citep[e.g.,][]{Eales09, RR14, Santini14, Mancini15, daCunha15, Nersesian19}. Although some works have extended dust measurements to include the CGM \citep{Menard10, PMC15}, such observations are rare. Therefore, as a first step, we have classified our model galaxies based on their cold dust content in the ISM. Figure~\ref{fig:dust-stellarmass} shows the relation between cold dust mass in the ISM and stellar mass of our model galaxies at redshift $z=0$. We find that the dust mass in the ISM increases with the host stellar mass, but with a significant scatter. 

The median values of our predicted ISM dust mass versus stellar mass are in a good agreement with the observations of dust in local galaxies, such as from the DustPedia catalogue \citep{Nersesian19}. Figure \ref{fig:dust-stellarmass} shows a large scatter below the median values, marking galaxies with relatively less dust compared to the overall population. 

We draw an arbitrary line at dust mass per stellar mass of $10^{-4}$, marked with the red line in Figure \ref{fig:dust-stellarmass} to divide the galaxy population into two categories: ISM dust-rich and ISM dust-poor. The ISM dust-poor group contains galaxies below the red line. Observations have found such ISM dust-poor galaxies in both the local and high-redshift Universe \citep{Fisher14}. 

Further exploration reveals that \dustysage galaxies in this ISM dust-poor category vary in their halo dust content. Therefore, we additionally divide the ISM dust-poor group based on their fraction of halo dust per stellar mass described below.

\subsubsection{Halo dust-rich}
\label{sssec:halo-dust-rich}

Several observations \citep{PMC15, Peeples14, Menard10} extended their search for metals and dust to the CGM. \citet{Peeples14} find that only $25\%$ of metals created in stars stay in the ISM; a similar fate also occurs for dust. \citet{Menard10} found galaxies with massive CGM dust. These observations discovered the existence of dust out of the ISM, which is reproduced by our model. 

In our model, a mix of halo and ejected dust might be a better representation of the CGM for some galaxies. However, we only use halo dust to represent the observed CGM dust. We use the same threshold as the ISM dust-mass to define the halo-dust rich galaxies: halo dust mass per stellar mass of $10^{-4}$. Halo dust-rich galaxies are defined as ISM dust-poor galaxies with halo dust mass above this threshold. 

\subsubsection{Dust-poor}
\label{sssec:dust-poor}

The population of galaxies that lack dust in both their ISM and halo is classified as dust-poor. Due to the rarity of dust measurements outside the ISM, we found no counterpart of this category from the observations. Their existence in our model serves as a prediction for future surveys measuring the CGM dust. 

\begin{figure}
    \centering
    \includegraphics[width=0.5\textwidth]{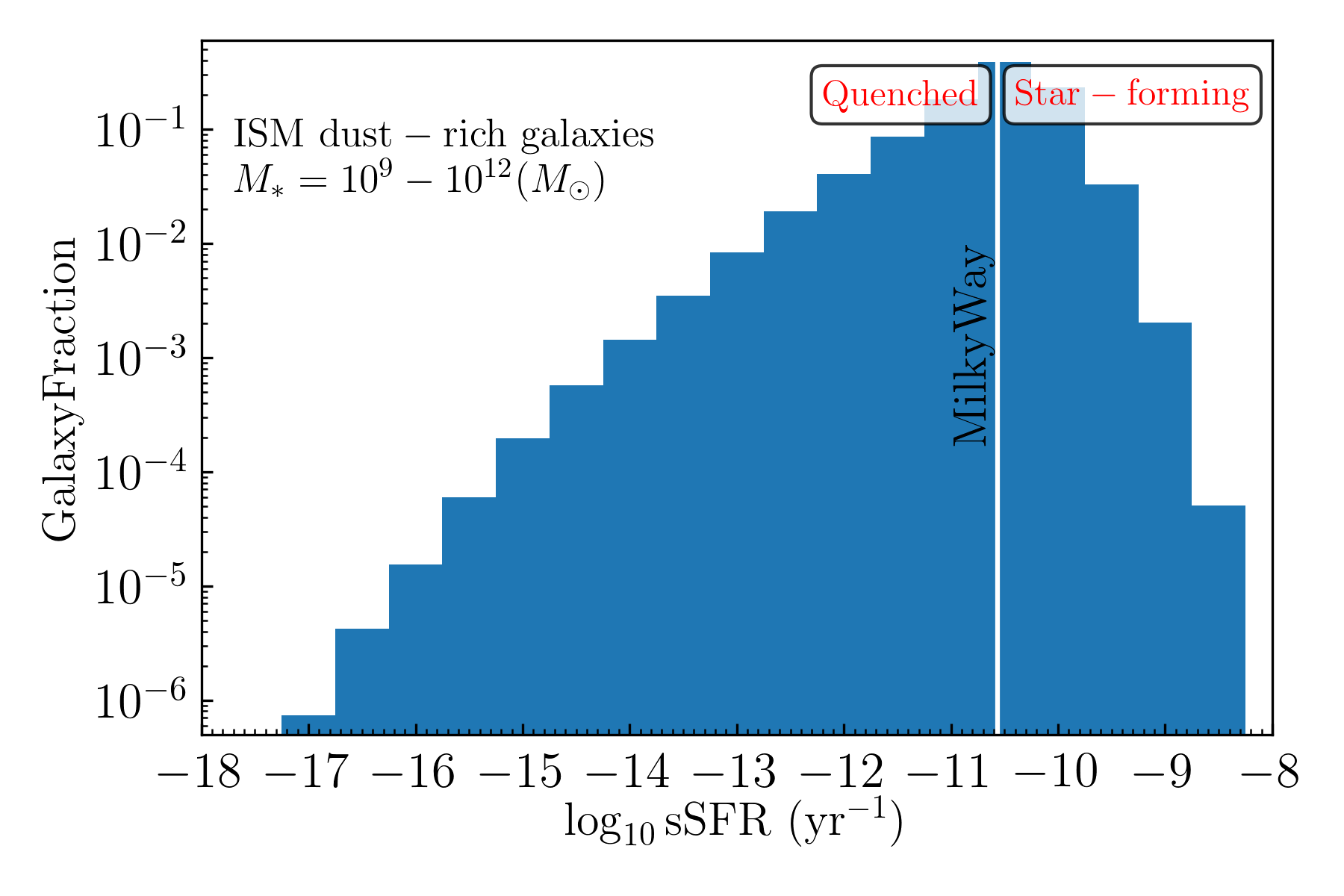}
    \caption{The distribution of the specific star-formation rate in our ISM dust-rich galaxies. The grey line at  $2.71 \times 10^{-11} \mathrm{yr^{-1}}$ is the Milky Way value adopted from \citet{Licquia15} which we use as a threshold in defining quenched and star-forming galaxies in our model galaxies.}
    \label{fig:sfr-hist}
\end{figure}

\subsection{ISM dust-rich}
\label{ssec:ISM-dust-rich}

We define the ISM dust-rich galaxies in our model as those with a fraction of ISM dust mass per stellar mass above $10^{-4}$. Dust obscures the intrinsic stellar spectra of such galaxies and re-emits the radiation in the infrared \citep{WTC92, WG00}. Infrared and sub-millimeter surveys have found a substantial number of these galaxies, especially at low redshift \citep[e.g.,][]{Eales09, DEE03, daCunha15, Clemens13, RR14}. 

Dust accounting in the galactic ISM is the main focus of many current and future infrared surveys. Forthcoming galaxy survey, like those using JWST, will measure dust in high redshift galaxies. Such surveys will provide additional constraints to our model predictions. 

Besides the relation between dust mass and stellar mass shown in Figure \ref{fig:dust-stellarmass}, observations have also found a relation between dust mass and SFR \citep{daCunha10, Casey12, Santini14}. Based on this, we divide our ISM dust-rich galaxies further into star-forming and quenched subclasses.

\subsubsection{Star-forming ISM dust-rich}
\label{sssec:sf-dust-rich}

Figure \ref{fig:sfr-hist} shows the specific SFR (sSFR) distribution of the ISM dust-rich galaxies in our model. To separate star-forming from quenched galaxies, we use the Milky Way sSFR value as the threshold, $2.71 \times 10^{-11} \mathrm{yr^{-1}}$ \citep{Licquia15}. Galaxies with sSFR above the Milky Way value are defined as the star-forming ISM dust-rich.

\subsubsection{Quenched ISM dust-rich}
\label{sssec:q-dust-rich}

ISM dust-rich galaxies are commonly related to star-forming spirals since dust is initially formed in stellar ejecta and grows in molecular clouds \citep{Valiante09}. However, we also find a population of ISM dust-rich galaxies with relatively low star formation activity in our model (see Figure \ref{fig:sfr-hist}). The quenched ISM dust-rich galaxies are those with a sSFR below the Milky Way value of  $2.71 \times 10^{-11} \mathrm{yr^{-1}}$.

\section{The relation between dust mass and fundamental galaxy properties}
\label{sec:relation}

\begin{figure*}
    \centering
    \includegraphics[width=1.0\textwidth]{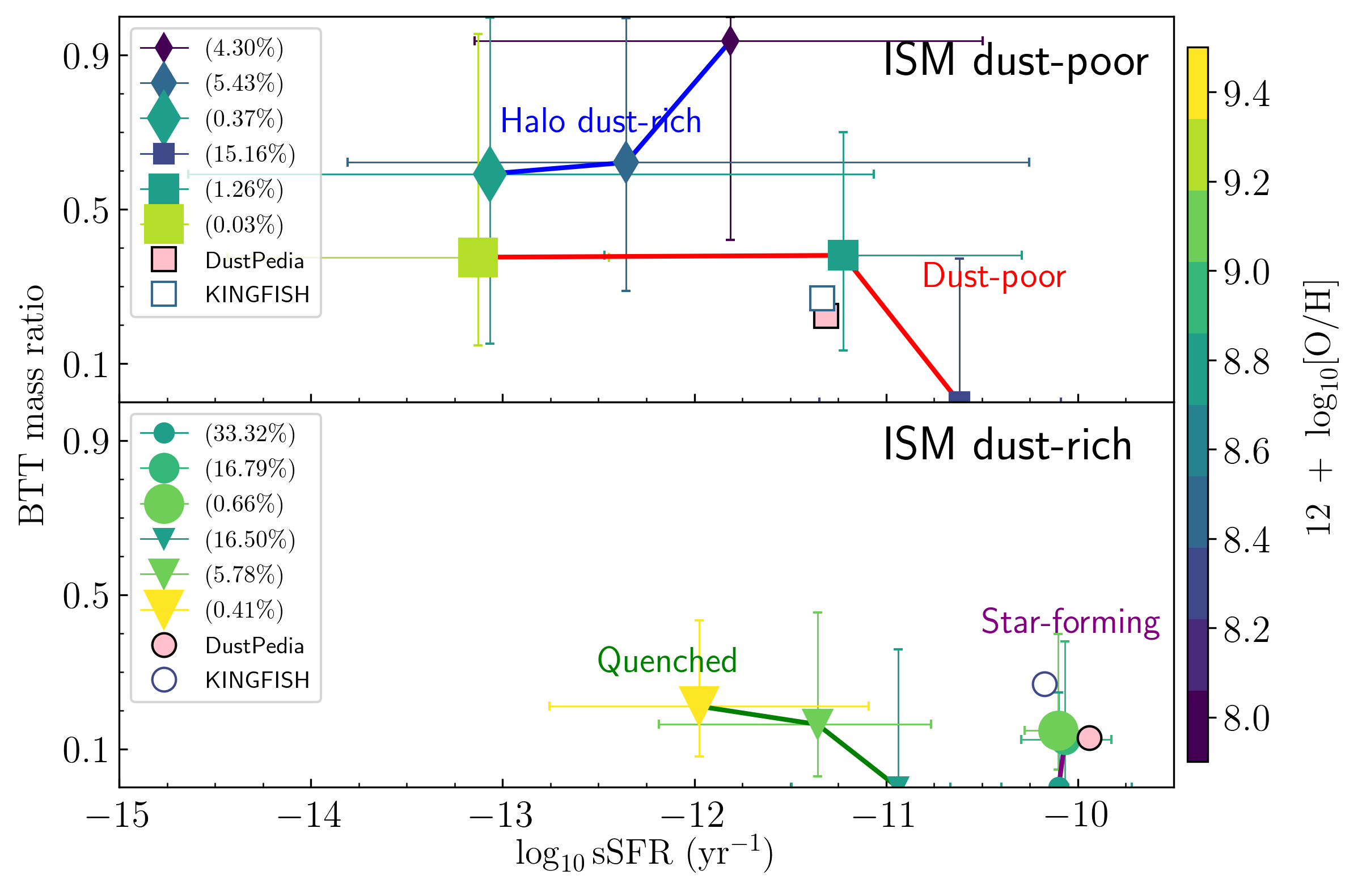}
    \caption{\textit{Top panel} shows the BTT ratio, sSFR and gas-phase metallicity of our model galaxies in the ISM dust-poor group. This group consists of two subgroups: halo dust-rich and dust-poor. Markers with  small, medium and large size represents three mass bin: $\log \mathrm{M_*} = 9 - 10$ \Msun, $\log \mathrm{M_*} = 10 - 11$, and $\log \mathrm{M_*} = 11 - 12$ \Msun, respectively. The points mark the median sSFR and median BTT mass ratio of each group, with the errorbars spanning the $16^\mathrm{th}$ and $84^\mathrm{th}$ percentiles. The colour of each point marks the gas-phase metallicity, with brighter colours representing higher value. We assume a solar metallicity of 0.02 and $12 + \log[\mathrm{O/H}] - \log[\mathrm{Z/Z_\odot}] = 9.0$. The number inside parentheses in the legend represents the percentage of galaxies in each category with respect to the total number of galaxies with stellar mass $\log \mathrm{M_*} = 9 - 12$ at $z=0$. Open symbols indicate equivalent observational dataset taken from DustPedia \citep{Nersesian19} and KINGFISH \citep{Kennicutt11, RR14}. 
    \textit{Bottom panel} shows the same relation above but for the ISM dust-rich group, which is splitted into star-forming and quenched subgroups.}
    \label{fig:btt-sfr}
\end{figure*}

In the previous sections, we categorized model galaxies into four classes: star-forming ISM dust-rich, quenched ISM dust-rich, halo dust-rich and dust-poor. We now plot the BTT mass ratio vs sSFR of each group in three mass bin in Figure \ref{fig:btt-sfr} as indicated in the legend, with the colour distribution representing the median gas-phase metallicity of each group. The error bars denote the $16^\mathrm{th}$ and $84^\mathrm{th}$ percentile range. 

Figure~\ref{fig:btt-sfr} shows a general trend where the ISM dust-rich galaxies in the bottom panel occupy the lower right space marking disky galaxies with higher star formation activity. Meanwhile, ISM dust-poor galaxies in the top panel generally have lower sSFR and a larger BTT mass ratio. Since galaxies typically change from star-forming spirals to quenched ellipticals over time, this plot indicates that the ISM also evolves from being dust-rich to dust-poor across h. 

The gas-phase metallicity in Figure~\ref{fig:btt-sfr} increase with stellar mass, as one may expect from the galaxy mass-metallicity relation \citep{Tremonti04}. Both ISM dust-rich groups generally have higher metallicity than the ISM dust-poor groups. 

Star-forming ISM dust-rich galaxies have the highest sSFR and lowest BTT mass ratio. Condensation in stellar ejecta is one of the main dust production channels in our model and is tightly correlated with star formation activity. Thus, we would expect ISM dust-rich galaxies to have a high sSFR. It is also possible that the high sSFR of such galaxies reflects the abundance of molecular gas in the ISM, which enables dust growth via accretion of metal grains.

Quenched ISM dust-rich galaxies have much lower sSFR (on average) and are also located in the low BTT-ratio regime. Their gas-phase metallicity is higher than that of the star-forming ISM dust-rich galaxies. As galaxies in this group have less sSFR than those in the star-forming group, their stellar dust production must be less efficient. However, their abundance of dust can be explained by their high availability of gas-phase metals, which can accrete onto existing dust grains in the dense environment. The low sSFR of these galaxies also suggest that they lack massive young stars that become SN, the primary destroyer of dust in the ISM. The scarcity of SN can also prevent the dust and refractory elements from being expelled, allowing them to preserve more dust and metals in their ISM. 

\begin{figure}
    \centering
    \includegraphics[width=0.5\textwidth]{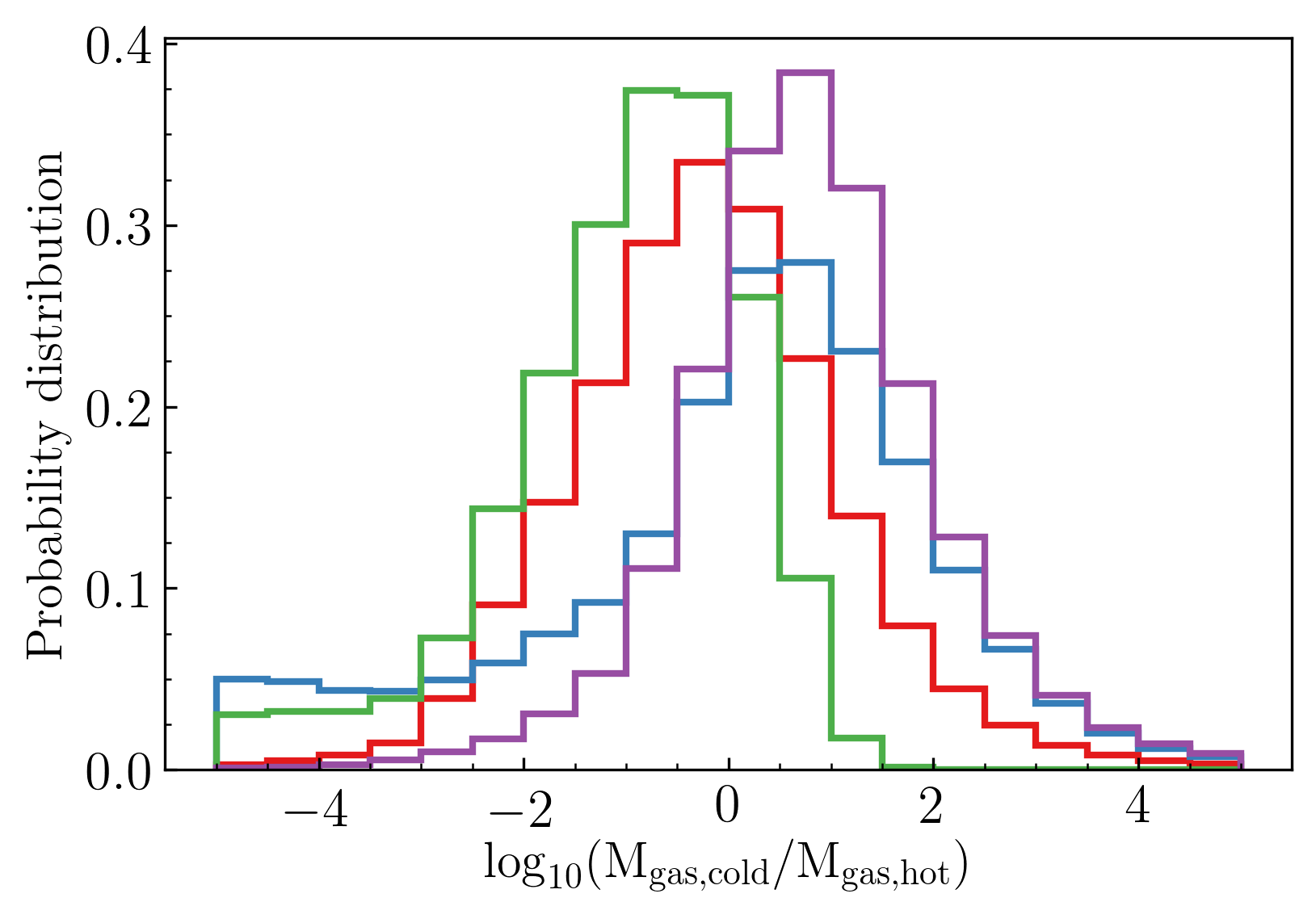}
    \caption{The probability distribution of the cold to hot gas mass ratio of our simulated galaxy categories. Red, blue, green and purple lines represent the star-forming ISM dust-rich, quenched ISM dust-rich, halo dust-rich and dust-poor groups, respectively.}
    \label{fig:hot-cold-ratio}
\end{figure}

In contrast to the ISM dust-rich galaxies, the ISM dust-poor group have higher BTT mass ratios (i.e., more elliptical), lower average sSFRs and lower gas-phase metallicities. The halo dust-rich group have the highest BTT mass ratio, incorporating the ``bulgiest'' galaxies for all mass bins. Galaxies in this group also have the lowest gas metallicities. To explore this further, we plot the distribution of the ratio between the ISM cold gas mass and the hot halo gas mass in Figure~\ref{fig:hot-cold-ratio}. Here, the halo dust-rich group peaks lowest of all groups, showing that galaxies in this group either have less cold gas in their ISM or more gas in their halo. The fact that these galaxies have massive halo dust and hot gas, yet they lack ISM gas and dust, indicates an efficient mechanism to transport their ISM content out of the disk.

\begin{figure}
    \centering
    \includegraphics[width=0.5\textwidth]{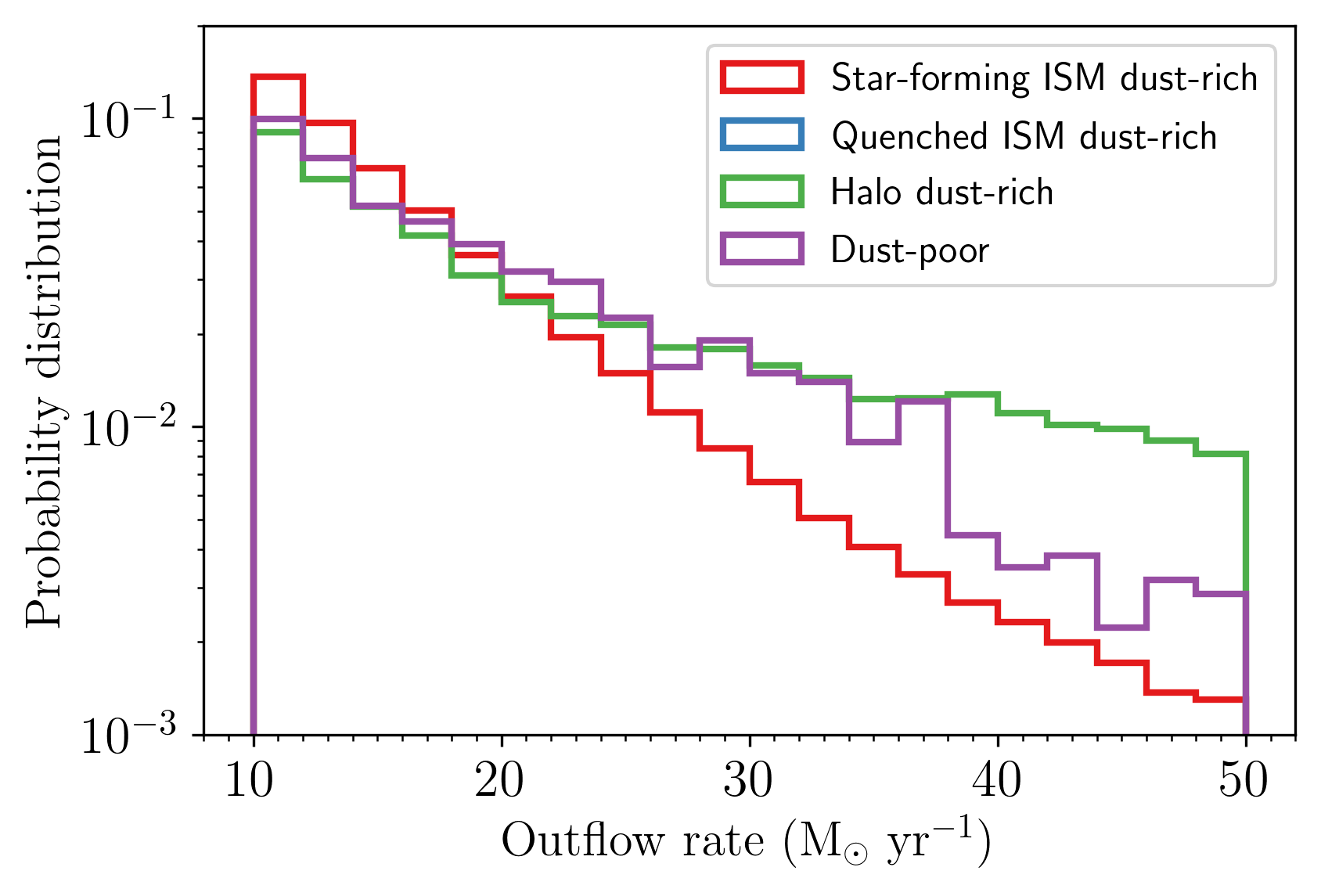}
    \caption{The gas outflow rate distribution due to feedback. We assume that the dust-to-gas mass ratio in the outflow is the same as the general ISM.}
    \label{fig:outflowrate}
\end{figure}

Figure~\ref{fig:outflowrate} shows the fraction of galaxies in each group versus outflow rate. The halo dust-rich group dominates for outflow rates above $30\ \Msun \mathrm{yr^{-1}}$. This suggests that the efficient outflow in these galaxies can blow their ISM content out of the disk, while their low sSFR does not allow them to replenish the metals and dust in the ISM. 

Looking back to Figure \ref{fig:btt-sfr}, we see that the total percentage of galaxies in the most massive mass bin ($\log \mathrm{M_*} = 11 - 12$) is $1.47 \%$ and the halo dust-rich group makes up a quarter of them, a more substantial contribution compared to those in the lower mass bins. In \dustysage, massive galaxies are likely to have deeper potential wells, which allow them to capture most of the dust outflow from the ISM in the halo. Therefore, very little dust escapes into the ejected reservoir in feedback events. Figure~\ref{fig:btt-fraction} also shows that in the most massive galaxies, the majority of the dust mass is in the hot halo, and an only small fraction of their dust fraction is in the ejected reservoir at all BTT mass ratios. 

In the dust-poor category, we find an anti-correlation between stellar mass and sSFR. The median sSFR of dwarf galaxies is $\log \mathrm{sSFR}=-10.6$, more than two orders of magnitude larger than those in the most massive bin. Galaxies will become dust-poor if their dust depletion rate exceeds the dust production rate. Since Figure \ref{fig:dust-stellarmass} shows a positive correlation between dust mass and stellar mass, more massive galaxies are more likely to have more dust in their ISM. Therefore, only those galaxies with a very low production rate can become dust-poor. The anti-correlation reflects the link between dust and star formation rates.

Figure \ref{fig:btt-sfr} also allows us to examine the stellar mass evolution across the four groups. At the massive end, galaxies with mass $\log \mathrm{M_*}\ (\Msun) = 11 - 12$ are comprised primarily of the star-forming ISM dust-rich category (0.6 \% of the total population), followed by the quenched ISM dust-rich and halo dust-rich groups each containing 0.3\% of the total galaxies. The dust-poor group is in the minority amongst massive galaxies. The dominance of the ISM dust-rich and halo dust-rich groups shows that the most massive galaxies rarely destroy or eject their dust out of the halo. Massive late-type and early-type galaxies contain the majority of their dust in the ISM and the halo, respectively.

In the middle bin ($\log \mathrm{M_*}\ (\Msun) = 10 - 11$), the biggest percentage is still contributed by the star-forming ISM dust-rich category (14 \%), followed by the quenched ISM dust-rich with 4.9 \%, halo dust-rich with 4.6 \%, and the dust-poor group with 1.1 \%. In the lowest mass bin ($\log \mathrm{M_*}\ (\Msun) = 9 - 10$), the star-forming ISM dust-rich still dominates (28\%). However, the contributions of quenched ISM dust-rich and dust-poor groups in this mass bin are also the most significant (13.9 \% and 12.8 \%, respectively) compared to those in the higher mass bins. The halo dust-rich group shares the smallest percentage for dwarf galaxies with only 3.6\%.

\subsection{Comparison with observed datasets}
\label{ssec:compared-with-data}

To add context to our theoretical predictions, we make comparisons with two observation datasets. DustPedia \citep{Nersesian19} is a compilation of local galaxies with dust measurements. To create this, the authors use the dust model \texttt{THEMIS} in the SED fitting code \texttt{CIGALE} to infer dust mass and galaxy properties from  \textit{Herschel} multiwavelength photometry. The authors use numerical Hubble T as the morphology indicator, with lower value corresponds to a larger stellar composition in bulge relative to the disk. To convert their Hubble T value to BTT-ratio, we use the K-band bulge-to-disk flux ratio mapping from \citet{Graham08}. Note that the BTT ratio for our simulated galaxies is based on stellar mass, while the observations are based on flux.

We group their data based on our definition of the star-forming ISM dust-rich,  quenched ISM dust-rich and ISM dust-poor galaxies. We can not divide their ISM dust-poor galaxies further because they do not provide a halo dust mass. We did not find galaxies classified as quenched ISM dust-rich in their dataset. Therefore, we plot the mean values of BTT mass ratio and sSFR of their star-forming ISM dust-rich group as a pink-filled circle and their ISM dust-poor group as a pink-filled square in Figure \ref{fig:btt-sfr}.

Our second observational dataset is taken from the KINGFISH survey \citep{Kennicutt11}. We group KINGFISH galaxies based on our definitions and compute the mean dust mass, BTT mass ratio, and SFR for each group. We plot the mean dust and galaxy properties with markers showing the group they belong. The dust measurement for KINGFISH galaxies is taken from \citet{RR14}.

Our prediction for star-forming ISM dust-rich and ISM dust-poor galaxies are in good agreement with the observed values. In both the KINGFISH and DustPedia data, dust-poor galaxies are those with the most elliptical morphology (E-type) with Hubble T below -2 \citep{RR14, Nersesian19}. However, the map of \citet{Graham08} that we use to convert the Hubble type to BTT ratio only extends to S0-type galaxies (Hubble T = -2). Therefore, we use the Hubble T = -2 conversion for elliptical galaxies with Hubble T < -2. The converted BTT flux ratio of these galaxies are $0.15$ dex lower than our median mass ratio but are still within our $16^\mathrm{th}$ percentile. 

\section{Evolution of the relations between dust and fundamental properties}
\label{sec:evolution}

\begin{figure*}
    \centering
    \includegraphics[width=1.0\textwidth]{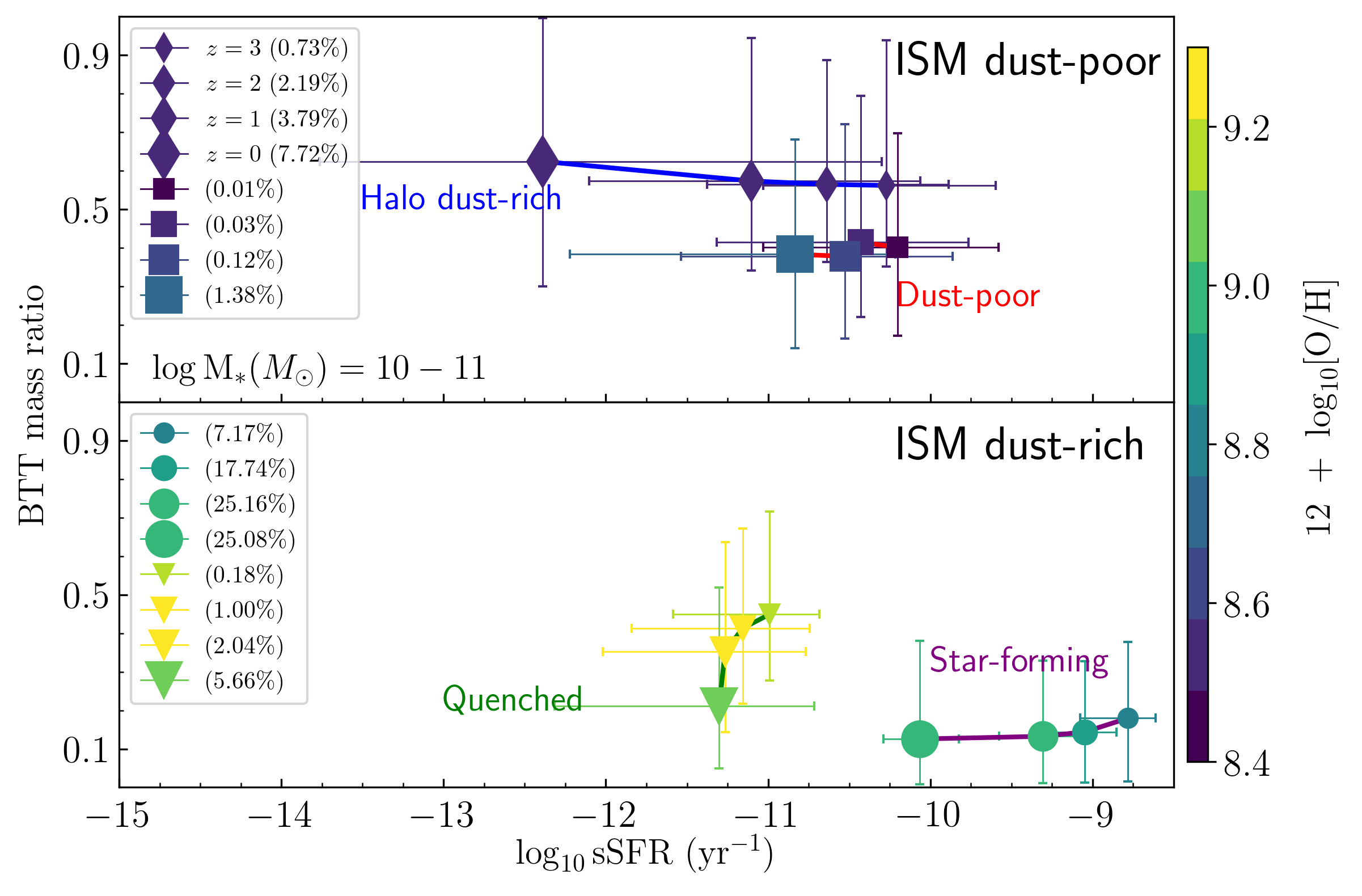}
    \caption{The evolution of the relation shown in Figure \ref{fig:btt-sfr} between redshift $z=0$ and redshift $z=3$. For clarity, we focus on the middle mass bin ($\log M* = 10 - 11$ \Msun). Circle, inverted triangle, diamond and square represent the median value of the star-forming ISM dust-rich, quenched ISM dust-rich, halo dust-rich and dust-poor groups, respectively.}
    \label{fig:btt-sfr-evol}
\end{figure*}

Galaxy evolution models predict that galaxies evolve from star-forming spiral to quench elliptical (i.e., the bottom right to top left in Figure \ref{fig:btt-sfr}) \citep{Somerville01, Baugh05, Croton06, Croton16}. The dust content of the ISM follows this evolutionary timeline to become dust-poor over time. There are two primary depletion mechanisms for ISM dust: destruction by SN shocks \citep{Slavin15} and ejection by SN and AGN feedback \citep{Feldmann15, Popping17}. Ejected dust can be trapped in the halo if the potential well is sufficiently deep, causing the host to end up as a halo dust-rich galaxy. Otherwise, such galaxies end up as dust-poor.

To see the evolution of the dust content in galaxies across time, we plot the relation between BTT mass ratio and sSFR of our dust groups from redshift $z=0$ to redshift $z=3$ in Figure \ref{fig:btt-sfr-evol}. Here, we focus on the middle mass bin ($\log \mathrm{M_*}\ (\Msun) = 10 - 11$) as representative of the Milky Way mass range. Circle, inverted triangle, diamond, and squares represent the star-forming ISM dust-rich, quenched ISM dust-rich, halo dust-rich, and dust-poor categories, respectively. This plot does not show the evolution of the same group of galaxies. Instead, it shows us the map of galaxies belong to each group at different redshift in the BTT mass ratio and sSFR plane.

Figure~\ref{fig:btt-sfr-evol} shows how the percentage of galaxies in each group changes over time. The fraction of star-forming ISM dust-rich galaxies peaks at $z=1$ then remains at $\approx 25\%$ until $z=0$. The other three groups show lower fractions but continue to increase down to the present day. This evolutionary trend reveals how star-forming ISM dust-rich galaxies might convert into the other groups at lower redshift. 

Across redshift, the behaviour of individual groups within the BTT mass ratio - sSFR space also changes. In the BTT mass ratio dimension, the quenched ISM dust-rich group gets diskier towards $z=0$ with a $0.3$ drop, while the other three groups do not show notable changes between $z=3$ and $z=0$. All groups except the quenched ISM dust-rich shows a significant decrease of the median sSFR from $z=3$ to $z=0$. The biggest sSFR decrease occurs in halo dust-rich galaxies, by more than 2 order of magnitude, while the star-forming ISM dust-rich and dust-poor galaxies each undergo a single dex decrease.

In the gas phase metallicity dimension, the halo dust-rich group have the lowest median value across redshift $z=3$ to $z=0$. The median metallicity of both dust-poor and star-forming ISM dust-rich galaxies increases slightly with decreasing redshift. Meanwhile, the median metallicity of the quenched ISM dust-rich group fluctuates with redshift; it increases $\approx 0.1$ dex between $z=3$ and $z=2$, does not change between $z=2$ and $z=1$, then decreases $\approx 0.3$ dex.

\section{Discussion: the behaviour of each group across redshift}
\label{sec:discussion}

From Figure~\ref{fig:btt-sfr} and Figure~\ref{fig:btt-sfr-evol}, we can see that although we divide galaxies into groups solely based on their dust content, the galaxy properties of each group are often distinct. Also, galaxy properties within each group evolve with redshift. Our predictions for how these groups change with time can help us understand the baryonic and dust physics occurring. In this section, we discussed the features of these groups between redshift $z=3$ and the present day.

\subsection{The star-forming ISM dust-rich group}

At all mass bins at redshift $z=0$, the star-forming ISM dust-rich group consistently stays at the lower right region in the BTT mass ratio - sSFR space (see Figure~\ref{fig:btt-sfr}). This indicates that these galaxies are still actively forming stars and have an abundance of molecular gas in their ISM, providing a dust production channel via condensation in stellar ejecta and grain growth in the molecular clouds. The gas-phase metallicity in this group is relatively high compared to those in the halo dust-rich and dust-poor groups. This group dominates the galaxy population for all mass bins at redshift $z=0$. 
    
Across redshift, this group stays in the disky star-forming region (see Figure~\ref{fig:btt-sfr-evol}) and continues to dominate in the overall population. The median gas-phase metallicity is lower at higher redshift. This metallicity increase with decreasing redshift reflects the ability of such galaxies always to retain an abundance of metals in their ISM. Since galaxies in the star-forming ISM dust-rich group have high sSFR by the selection, we expect that the ISM will be both metal and dust-rich. Observations of the local and high redshift Universe find many galaxies that belong to this group. Figure \ref{fig:btt-sfr} shows that at redshift $z=0$, both the KINGFISH~\citep{Kennicutt11, RR14} and DustPedia~\citep{Nersesian19} datasets have galaxies with dust-to-stellar mass and sSFR classifications similar to our star-forming ISM dust-rich group. 

\begin{figure*}
    \centering
    \includegraphics[width=1\textwidth]{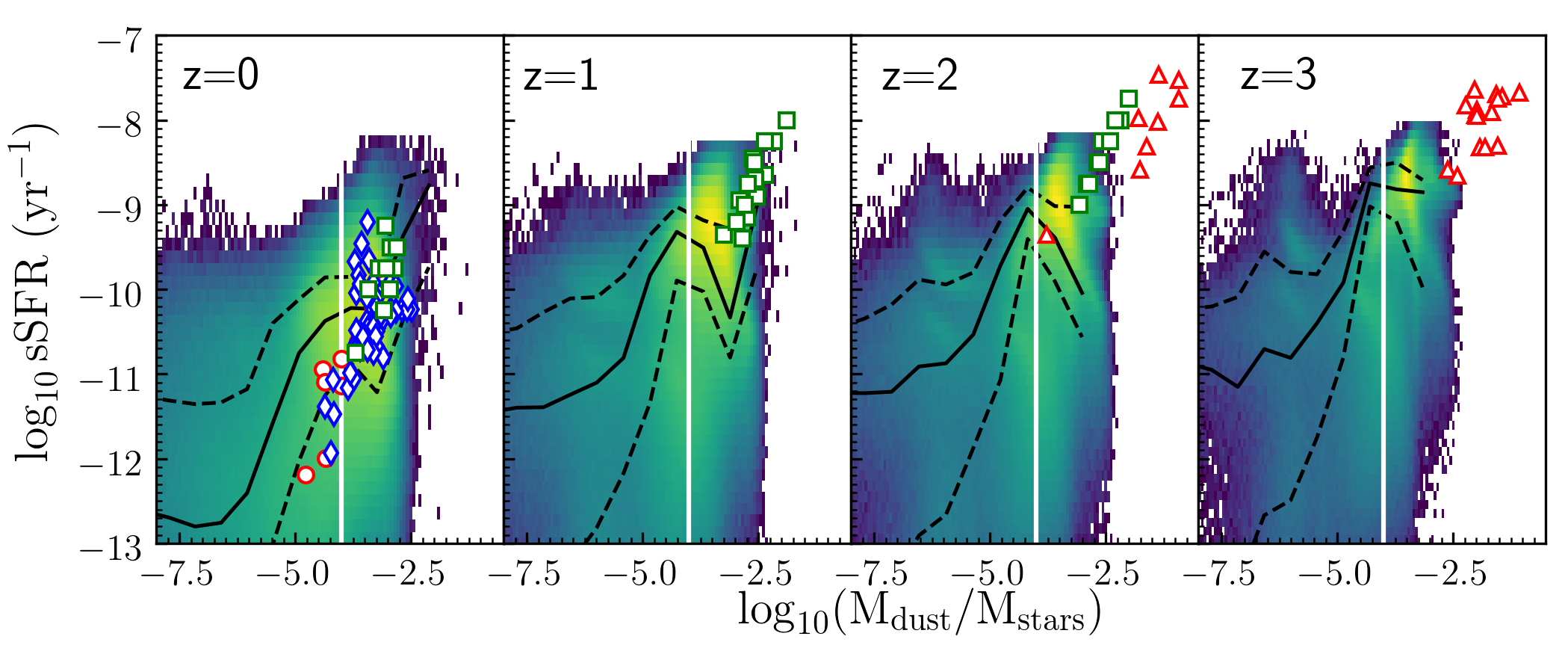}
    \caption{The relation between specific star formation rate (sSFR) and the dust-to-stellar mass ratio from redshift $z=0$ to $z=3$. The heat map shows the 2D density distribution of our model galaxies with brighter color representing higher density. The solid black lines mark the median while the dashed lines marked the $16^\mathrm{th}$ and $84^\mathrm{th}$ percentile. The white vertical lines mark a log dust-to-stellar mass ratio = -4, where we divide the ISM dust-rich and dust-poor categories. The red circles, blue diamonds, green squares and red triangles are observational values from DustPedia: \citet{Nersesian19}, \citet{RR14}, \citet{Santini14} and \citet{daCunha15}, respectively.}
    \label{fig:dtm-ssfr}
\end{figure*}

Figure \ref{fig:dtm-ssfr} shows the sSFR of our model galaxies as a function of dust-to-stellar mass ratio from redshift $z=0$ to $z=3$. The white line marks the divider between the ISM dust-rich and ISM dust-poor groups. Across redshift, many authors find star-forming galaxies with a dust rich ISM. The general trend that the sSFR decreases with decreasing redshift is seen both for the model and in the observations. At $z=0$, our ISM dust-rich galaxies are in good agreement with the observational values from DustPedia \citep{Nersesian19}, \citet{RR14} and \citet{Santini14}. For the ISM dust poor galaxies below the grey line, the observations are lower than our median value but are still within the $16^\mathrm{th}$ percentile. At higher redshift, observations are yet to find ISM dust-poor galaxies. At $z=1$, our model roughly agrees with the dust-rich data from \citet{Santini14}, although our median value of sSFR lies $0.1 - 0.4$ dex below the observed values. At $z=2$ and $z=3$, however, our model fails to reproduce galaxies with $\log (\mathrm{M_{dust}} / \mathrm{M_{star}}) > -2.5$ where most of the observational datasets lie. \dustysage and our base model \sage \citep{Croton16} are constrained using various galaxy observations at $z=0$; therefore, we still find limitations in reproducing the high redshift population.  We also note that the dataset from \citet{daCunha15} (red triangles) is obtained in the ALMA LESS survey which is biased towards the extremely bright galaxies.

\subsection{The quenched ISM dust-rich group}

The quenched ISM dust-rich group consists of galaxies that have an abundance of dust in their ISM despite their low star formation activity (defined as galaxies with sSFR below the Milky Way). One possible scenario for this is that due to the low sSFR, these galaxies no longer host supernovae events that destroy dust and expel metals and dust from the ISM. This also explains the higher gas-phase metallicity of galaxies in this group compared to those in the star-forming ISM dust-rich group. The abundance of metals in the ISM provides material for dust growth via grain accretion, allowing such galaxies to accumulate dust mass. Another scenario for their dust production is the condensation in the ejecta from previous star formation episodes. Although they do not have a fresh supply of stellar ejecta due to the low sSFR, it is possible that there are still refractory elements from previous SN and AGB winds in the ISM that condense into dust grains.

Figure~\ref{fig:btt-sfr} shows that the majority of quenched ISM dust-rich galaxies can be found in the low mass population. Compared to the star-forming ISM dust-rich group, this group is rare at low and high redshift. Figure~\ref{fig:btt-sfr-evol} shows that the fraction of this group with stellar mass $\log \mathrm{M_*} = 10-11$ at $z > 0$ is always below $6\%$. Due to their low abundance, we have not found any observational evidence for this group.
 
\subsection{The halo dust-rich group}

Halo dust-rich galaxies have little or no dust in their ISM but contain a significant amount of dust in their halo. At redshift $z=0$, this group includes elliptical galaxies with the largest BTT mass ratios and low sSFR. The median gas-phase metallicity of this group is also the lowest compared to the other groups. The small sSFR does not allow for efficient instantaneous metal and stellar dust production in such galaxies. However, they should have accumulated metals and dust from previous star-formation episodes. The lack of metals and dust in their ISM, while their halo is dust-rich, indicates that the ISM content is transported into the halo (see Section~\ref{sec:relation}).

At higher redshift, however, the halo dust-rich group occupies an increasingly high sSFR region with a median $\log \mathrm{sSFR} >$ the Milky Way value. The high sSFR of these galaxies means that they are producing metals and dust in their stellar ejecta. However, unlike galaxies in the other groups, galaxies in this group show no increase in gas metallicity with decreasing redshift. The low abundance of metals and dust in their ISM is possible if the feedback is efficiently heating and blowing out the newly formed metals and dust. Such feedback might be provided by AGN activity \citep{Sarangi19}.

Galaxies in a deep potential well can retain the feedback-heated gas, dust and metals in their halo. Mass in the dark matter halo usually sets the depth of the potential well. Since stellar mass correlates with halo mass, halo dust-rich galaxies should be more common in the massive galaxies. Figure \ref{fig:btt-sfr} shows that in the massive bin, this group comprise a quarter of the population, a more significant fraction than the lower stellar mass bin.

\citet{Menard10} measured dust reddening effects of background quasars relative to the foreground SDSS galaxies. They observed the existence of diffuse dust in halos with an amount comparable to those in the galactic disk. For $0.5L^*$ galaxies, the authors estimated a dust mass of $5 \times 10^7 \Msun$ in the halo. \citet{PMC15}
confirmed this halo dust mass for $0.1L^* - 1L^*$ low redshift galaxies. Observations of CGM dust provides evidence for our prediction that dust is transported out of the ISM and trapped in the halo. However, such observations are still uncommon. Future CGM dust measurements will allow for better constraints for models such as \dustysage.

\subsection{The dust-poor group}

We define galaxies that lack dust in both the ISM and their halo as dust-poor galaxies. At redshift $z=0$ (Figure~\ref{fig:btt-sfr}), the dust-poor group has low median sSFR, low gas-phase metallicity and high BTT mass ratio. In the DustPedia~\citep{Nersesian19} and KINGFISH~\citep{Kennicutt11, RR14} datasets, galaxies with morphological type E-S0 have a median dust-to-stellar mass ratio below $1 \times 10^{-4}$ and are classified as ISM dust-poor as per our definition. However, both datasets do not provide measurement for dust in the halo. Therefore, we assume their halo is dust-poor, and we include them in the dust-poor group instead of the halo dust-rich group.
  
Dust depletion in the ISM occurs in two ways: destruction by SN shocks inside the galaxy disk and dust that is ejected out of the ISM by feedback. If the potential well of the system is enough to trap the dust, the ejected dust will end up in halo. However, galaxies with low stellar mass usually live in shallower potentials and thus are likely to lose their dust completely through ejection. This likely explains why Figure~\ref{fig:btt-sfr} shows that the dust-poor group contain the most significant fraction of their galaxies in the low mass bin.

In Figure \ref{fig:btt-sfr-evol}, the dust-poor group moves to the lower BTT mass ratio, lower sSFR region with decreasing redshift while the gas metallicity increases. At high redshift, galaxies in this group have relatively high sSFR, but their ISM still lacks both metals and dust. Dust formation is highly sensitive to the abundance of metals. In stellar ejecta, dust forms via the condensation of metals ejected from stars. In dense clouds, dust grains grow by accreting metals. Therefore, galaxies with low metallicity can not produce dust efficiently. If this low production rate can not keep up with the depletion rate, a galaxy will end up in the dust-poor group.

\citet{Fisher14} observed the dust emission from a local dwarf galaxy, I Zwicky 18, and found it lacks dust with a dust-to-stellar mass ratio of about $10^{-5}$ to $10^{-6}$. This galaxy has a SFR of $0.05$ \Msun/year and a very low metallicity of $12+\log\mathrm{[O/H]} = 7.17$, much lower than that seen in local galaxies \citep{Fisher14}. The reported stellar mass is $9 \times 10^7 \Msun$, which makes the sSFR $ = 5.5 \times 10^{-10}$, consistent with the median sSFR value of the dust-poor galaxies in the lower mass group from \dustysage (see Figure \ref{fig:btt-sfr}). Although the stellar mass of I Zwicky 18 is well below our sample, we find a consistent feature where a lack of metals accompanies the lack of dust. 

\section{Conclusion}
\label{sec:conclusion}

During galaxy's evolution, dust physics is heavily influenced by star formation, gas physics, outflows and the depth of the potential well. Therefore, the dust, stellar, and gas content of galaxies are tightly related to each other. In this paper, we use the \dustysage galaxy formation model \mbox{\citep{Triani20}} to explore the relation between dust content and the fundamental properties of galaxies. We have divided the galaxy population based on their dust mass and observable properties. At redshift $z=0$, the properties of our model galaxies are in good agreement with observations, especially where the statistics are sound. 

Our investigation includes the comparison of morphology, specific star formation rate, metallicity and stellar mass for galaxies grouped by their dust content from redshift $z=3$ to $z=0$. Our main conclusions are the following:

\begin{itemize}
    \item We find that the distribution of dust in the ISM, hot halo and ejected reservoir is affected by galaxies' morphology and stellar mass. The fraction of dust in the hot halo increases with BTT mass ratio and stellar mass (Figure \ref{fig:btt-fraction}).
    
    \item At redshift $z=0$, ISM dust-rich galaxies have the highest gas-phase metallicities. Both star-forming and quenched ISM dust-rich groups shows a low BTT mass ratio and relatively high median sSFR. Galaxies in the halo dust-rich and dust-poor groups are metal poor. They are distributed in the high BTT mass ratio and low sSFR regime of Figure~\ref{fig:btt-sfr}.
    
    \item We see an evolution in the behaviour of our grouped galaxies in Figure~\ref{fig:btt-sfr-evol}. From $z=3$ to $z=0$: (i) the BTT mass ratio of the quenched ISM dust-rich group drops $\approx 0.3$ dex while the other three groups show insignificant changes, (ii) the median sSFR of all groups except quenched ISM dust-rich trend towards lower region, and (iii) all groups show weak evolution of gas-phase metallicity with redshift except the quenched ISM dust-rich group where the metallicity fluctuates with redshift.
    
    \item Across redshift, both star-forming ISM dust-rich and quenched ISM dust-rich groups maintain their high median sSFR and high gas-phase metallicity. This implies that such galaxies are still actively producing metals and dust via stellar production. The abundance of metals in the ISM also enables dust growth via metal accretion in dense clouds.
    
    \item The halo dust-rich group consists of galaxies with a dust-poor ISM but dust-rich halo. They have a relatively high outflow rate of heated gas, dust, and metals out of the ISM. At present, they are elliptical with very low sSFR and metallicity. The gas-phase metallicity of these galaxies does not depend significantly on redshift. At high redshift, galaxies in this group have high sSFR that enables an efficient dust production mechanism. Their lack of ISM dust, therefore, implies effective feedback to reheat the newly formed dust. Because these massive galaxies have a deep potential well, they are more likely to retain the heated dust in their halo. Figure~\ref{fig:btt-sfr} and Figure \ref{fig:btt-sfr-evol} shows that this group is made up of a larger fraction of massive galaxies.
    
    \item Galaxies in the dust-poor group have low dust mass both in their ISM and halo. From redshift $z=3$ to $z=0$, this group evolves from the high sSFR region to the lower sSFR region. However, they always have low metallicity. As metals are the main ingredients for dust, this low metallicity implies that their dust production is also small.
\end{itemize}

Our model provides predictions for future surveys with next-generation instruments and telescopes, such as ALMA and JWST, that will measure the dust and galaxy properties in extraordinary detail at high redshift. JWST will cover the IR spectrum in the 0.6 to 28.5 micron regime, which reveals the total galactic dust content up to $z=2$. At $z=3$, it will only cover the dust spectrum to $\approx 7$ micron, providing a measure for the mass of hot dust. As we have seen throughout this paper, current dust measurement at high-redshift are only available for the brightest galaxies with high dust mass. Complemented with data from ALMA and other IR and sub-mm surveys, JWST will provide a significantly improved view of high redshift dust across a wider range of masses. This will better constraint dust enrichment scenarios in models such as \dustysage, leading to a better predictions and more detailed interpretation of the observations. 

\section*{Acknowledgements}
We would like to thank the anonymous referee whose valuable comments improved the quality of this paper, and Ned Taylor for helpful comments during the final stages of this work. This research was supported by the Australian Research Council Centre of Excellence for All Sky Astro-physics in 3 Dimensions (ASTRO 3D), through project number CE170100013.  The Semi-Analytic Galaxy Evolution (SAGE) model, on which \dustysage was built, is a publicly available codebase that runs on the dark matter halo trees of a cosmological N-body simulation. It is available for download at \url{https://github.com/darrencroton/sage}. This research has used \texttt{python} (\url{https://www.python.org/}), \texttt{numpy} \citep{Vanderwalt11} and \texttt{matplotlib} \citep{Hunter07}.

\section*{Data availability}
The data underlying this article are available in the article. The galaxy formation model used to generate the data is available at \url{https://github.com/dptriani/dusty-sage}.



\bibliographystyle{mnras}
\bibliography{bibliography} 








\bsp	
\label{lastpage}
\end{document}